\pdfoutput=1

\documentclass[12pt,a4paper]{article}

\usepackage{ifthen} 
\newboolean{pdflatex}
\setboolean{pdflatex}{true} 

\newboolean{articletitles}
\setboolean{articletitles}{true} 

\newboolean{uprightparticles}
\setboolean{uprightparticles}{false} 

\newboolean{inbibliography}
\setboolean{inbibliography}{false} 

\usepackage[top=1in, bottom=1.25in, left=1in, right=1in]{geometry}

\usepackage{microtype}
\usepackage{lineno}  
\usepackage{xspace} 
\usepackage{caption} 

\usepackage{graphicx}  
\usepackage{color}
\usepackage{colortbl}
\graphicspath{{./figs/}} 

\usepackage{amsmath} 
\usepackage{amssymb}
\usepackage{amsfonts}
\usepackage{upgreek} 

\newcommand*\patchAmsMathEnvironmentForLineno[1]{%
\expandafter\let\csname old#1\expandafter\endcsname\csname #1\endcsname
\expandafter\let\csname oldend#1\expandafter\endcsname\csname
end#1\endcsname
 \renewenvironment{#1}%
   {\linenomath\csname old#1\endcsname}%
   {\csname oldend#1\endcsname\endlinenomath}%
}
\newcommand*\patchBothAmsMathEnvironmentsForLineno[1]{%
  \patchAmsMathEnvironmentForLineno{#1}%
  \patchAmsMathEnvironmentForLineno{#1*}%
}
\AtBeginDocument{%
\patchBothAmsMathEnvironmentsForLineno{equation}%
\patchBothAmsMathEnvironmentsForLineno{align}%
\patchBothAmsMathEnvironmentsForLineno{flalign}%
\patchBothAmsMathEnvironmentsForLineno{alignat}%
\patchBothAmsMathEnvironmentsForLineno{gather}%
\patchBothAmsMathEnvironmentsForLineno{multline}%
\patchBothAmsMathEnvironmentsForLineno{eqnarray}%
}

\usepackage{hyperref}    
\usepackage[all]{hypcap} 

\usepackage{xspace} 
\usepackage{upgreek}

\def\lhcb {\mbox{LHCb}\xspace}

\def\MagUp {\mbox{\em Mag\kern -0.05em Up}\xspace}

\ifthenelse{\boolean{uprightparticles}}%
{

 \def\PDelta      {\ensuremath{\Delta}\xspace}                 
 \def\PXi      {\ensuremath{\Xi}\xspace}                 
 \def\PLambda      {\ensuremath{\Lambda}\xspace}                 
 \def\PSigma      {\ensuremath{\Sigma}\xspace}                 
 \def\POmega      {\ensuremath{\Omega}\xspace}                 
 \def\PUpsilon      {\ensuremath{\Upsilon}\xspace}                 
 
 \def\PB      {\ensuremath{\mathrm{B}}\xspace}                 
                  
 \def\PD      {\ensuremath{\mathrm{D}}\xspace}

 \def\PK      {\ensuremath{\mathrm{K}}\xspace}

 \def\Pb      {\ensuremath{\mathrm{b}}\xspace}                 
 \def\Pc      {\ensuremath{\mathrm{c}}\xspace}

 \def\Pi      {\ensuremath{\mathrm{i}}\xspace}

 \def\Pp      {\ensuremath{\mathrm{p}}\xspace}

}
{

 \mathchardef\PDelta="7101
 \mathchardef\PXi="7104
 \mathchardef\PLambda="7103
 \mathchardef\PSigma="7106
 \mathchardef\POmega="710A
 \mathchardef\PUpsilon="7107
                  
 \def\PB      {\ensuremath{B}\xspace}                 
                  
 \def\PD      {\ensuremath{D}\xspace}

 \def\PK      {\ensuremath{K}\xspace}

 \def\Pb      {\ensuremath{b}\xspace}                 
 \def\Pc      {\ensuremath{c}\xspace}

 \def\Pi      {\ensuremath{i}\xspace}

 \def\Pp      {\ensuremath{p}\xspace}

}

\makeatletter
\ifcase \@ptsize \relax
  \newcommand{\miniscule}{\@setfontsize\miniscule{4}{5}}
\or
  \newcommand{\miniscule}{\@setfontsize\miniscule{5}{6}}
\or
  \newcommand{\miniscule}{\@setfontsize\miniscule{5}{6}}
\fi
\makeatother

\DeclareRobustCommand{\optbar}[1]{\shortstack{{\miniscule (\rule[.5ex]{1.25em}{.18mm})}
  \\ [-.7ex] $#1$}}

\def\cquark    {{\ensuremath{\Pc}}\xspace}

\def\bquark    {{\ensuremath{\Pb}}\xspace}

  \def\Kbar    {{\kern 0.2em\overline{\kern -0.2em \PK}{}}\xspace}

\def\KorKbar    {\kern 0.18em\optbar{\kern -0.18em K}{}\xspace}

  \def\Dbar    {{\kern 0.2em\overline{\kern -0.2em \PD}{}}\xspace}

\def\DorDbar    {\kern 0.18em\optbar{\kern -0.18em D}{}\xspace}

\def\Bbar    {{\ensuremath{\kern 0.18em\overline{\kern -0.18em \PB}{}}}\xspace}

\def\BorBbar    {\kern 0.18em\optbar{\kern -0.18em B}{}\xspace}

  \def\Y#1S{\ensuremath{\PUpsilon{(#1S)}}\xspace}

\def\proton      {{\ensuremath{\Pp}}\xspace}

\def\Lbar        {{\ensuremath{\kern 0.1em\overline{\kern -0.1em\PLambda}}}\xspace}
\def\LorLbar    {\kern 0.18em\optbar{\kern -0.18em \PLambda}{}\xspace}

\def\eps   {{\ensuremath{\varepsilon}}\xspace}

\def\AT#1     {\ensuremath{A_{\mathrm{T}}^{#1}}\xspace}           

\def\C#1      {\ensuremath{\mathcal{C}_{#1}}\xspace}                       
\def\Cp#1     {\ensuremath{\mathcal{C}_{#1}^{'}}\xspace}                    
\def\Ceff#1   {\ensuremath{\mathcal{C}_{#1}^{\mathrm{(eff)}}}\xspace}        
\def\Cpeff#1  {\ensuremath{\mathcal{C}_{#1}^{'\mathrm{(eff)}}}\xspace}       
\def\Ope#1    {\ensuremath{\mathcal{O}_{#1}}\xspace}                       
\def\Opep#1   {\ensuremath{\mathcal{O}_{#1}^{'}}\xspace}                    

\newcommand{\tev}{\ifthenelse{\boolean{inbibliography}}{\ensuremath{~T\kern -0.05em eV}\xspace}{\ensuremath{\mathrm{\,Te\kern -0.1em V}}}\xspace}
\newcommand{\gev}{\ensuremath{\mathrm{\,Ge\kern -0.1em V}}\xspace}
\newcommand{\mev}{\ensuremath{\mathrm{\,Me\kern -0.1em V}}\xspace}
\newcommand{\kev}{\ensuremath{\mathrm{\,ke\kern -0.1em V}}\xspace}
\newcommand{\ev}{\ensuremath{\mathrm{\,e\kern -0.1em V}}\xspace}
\newcommand{\gevc}{\ensuremath{{\mathrm{\,Ge\kern -0.1em V\!/}c}}\xspace}
\newcommand{\mevc}{\ensuremath{{\mathrm{\,Me\kern -0.1em V\!/}c}}\xspace}
\newcommand{\gevcc}{\ensuremath{{\mathrm{\,Ge\kern -0.1em V\!/}c^2}}\xspace}
\newcommand{\gevgevcccc}{\ensuremath{{\mathrm{\,Ge\kern -0.1em V^2\!/}c^4}}\xspace}
\newcommand{\mevcc}{\ensuremath{{\mathrm{\,Me\kern -0.1em V\!/}c^2}}\xspace}

\def\mum  {\ensuremath{{\,\upmu\mathrm{m}}}\xspace}

\def\mbarn{\ensuremath{\mathrm{ \,mb}}\xspace}

\def\gsim{{~\raise.15em\hbox{$>$}\kern-.85em
          \lower.35em\hbox{$\sim$}~}\xspace}
\def\lsim{{~\raise.15em\hbox{$<$}\kern-.85em
          \lower.35em\hbox{$\sim$}~}\xspace}

\def\ptot       {\mbox{$p$}\xspace}
\def\pt         {\mbox{$p_{\mathrm{ T}}$}\xspace}

\newcommand{\lum} {\ensuremath{\mathcal{L}}\xspace}

\def\evtgen     {\mbox{\textsc{EvtGen}}\xspace}

\def\geant      {\mbox{\textsc{Geant4}}\xspace}

\def\photos     {\mbox{\textsc{Photos}}\xspace}

\def\pythia     {\mbox{\textsc{Pythia}}\xspace}

\def\tell1  {TELL1\xspace}
\def\ukl1   {UKL1\xspace}

\newcommand{\eg}{\mbox{\itshape e.g.}\xspace}
\newcommand{\ie}{\mbox{\itshape i.e.}\xspace}

\newcommand{\beq}[1]{\begin{equation}{\label{#1}}}
\newcommand{\eeq}[0]{\end{equation}}
\newcommand{\barr}[0]{\begin{array}}
\newcommand{\earr}[0]{\end{array}}

\usepackage{cite} 
\usepackage{mciteplus}

\usepackage{longtable} 

\usepackage{graphicx}
\usepackage{caption}
\usepackage{subcaption}

\usepackage{multirow}
\usepackage{graphicx}
\usepackage{lineno}
\usepackage{amsmath}
\usepackage{booktabs}
\usepackage{dcolumn}
\newcolumntype{.}{D{.}{.}{-1}}

\begin{document}

\renewcommand{\thefootnote}{\fnsymbol{footnote}}
\setcounter{footnote}{1}


\begin{titlepage}
\pagenumbering{roman}

\vspace*{-1.5cm}
\centerline{\large EUROPEAN ORGANIZATION FOR NUCLEAR RESEARCH (CERN)}
\vspace*{1.5cm}
\noindent
\begin{tabular*}{\linewidth}{lc@{\extracolsep{\fill}}r@{\extracolsep{0pt}}}
\ifthenelse{\boolean{pdflatex}}
{\vspace*{-3.0cm}\mbox{\!\!\!\includegraphics[width=.14\textwidth]{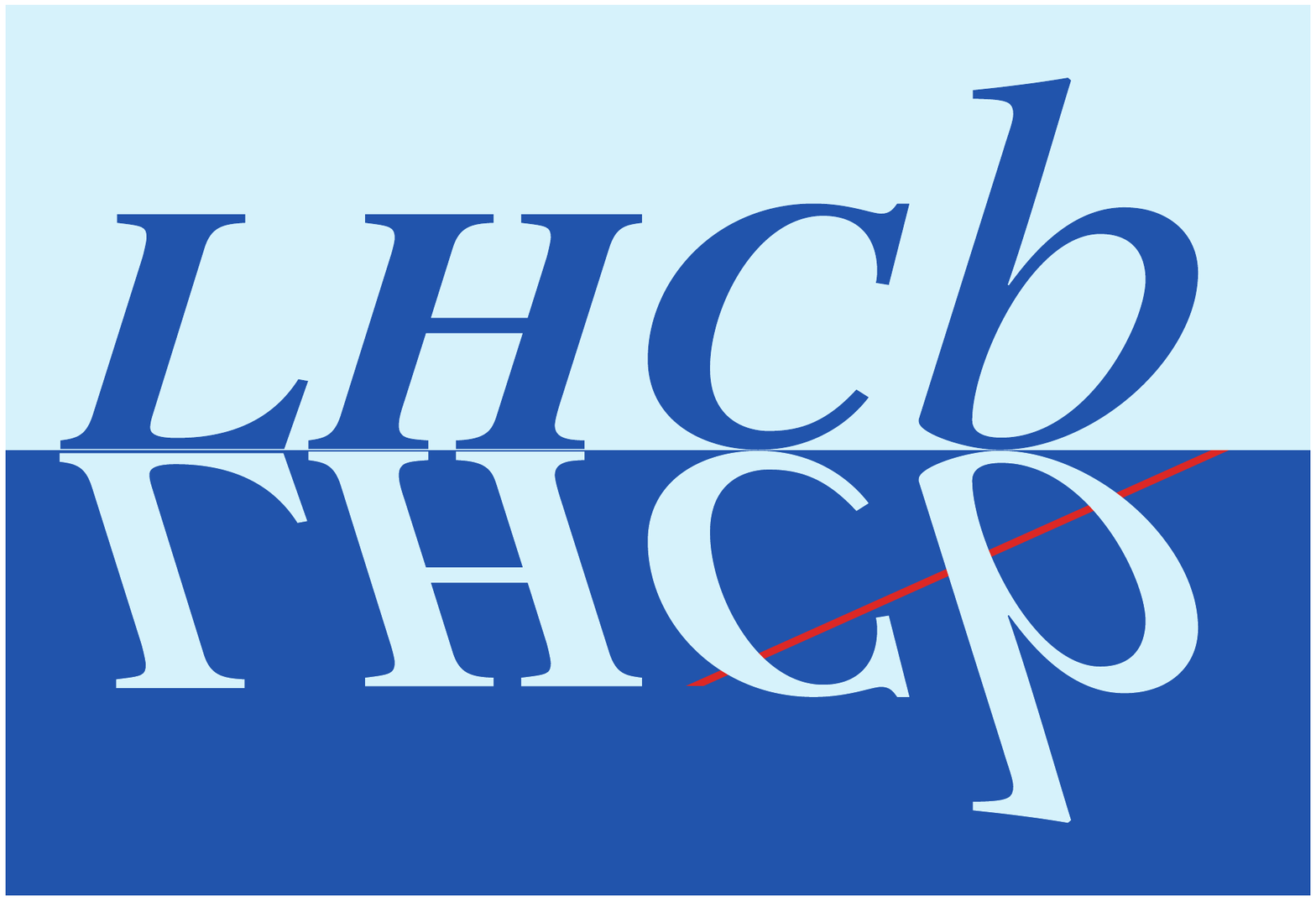}} & &}%
{\vspace*{-1.2cm}\mbox{\!\!\!\includegraphics[width=.12\textwidth]{lhcb-logo.eps}} & &}%
\\
 & & CERN-EP-2018-044 \\  
 & & LHCb-PAPER-2018-003 \\  
 & & June 10, 2018 \\ 
 & & \\ 
\end{tabular*}

\vspace*{2.0cm}

{\normalfont\bfseries\boldmath\huge
\begin{center}
Measurement of the inelastic \proton\proton cross-section 
at a centre-of-mass energy of 13\,TeV
\end{center}
}

\vspace*{1.0cm}

\begin{center}
The LHCb collaboration\footnote{Authors are listed at the end of this paper.}
\end{center}

\vspace*{1.0cm}

\begin{abstract}
  \noindent
The cross-section for inelastic proton-proton collisions at a centre-of-mass
energy of 13\,TeV is measured with the \lhcb\ detector. The fiducial 
cross-section for inelastic interactions producing at least one prompt long-lived 
charged particle with momentum $p>2$\,\gevc in the pseudorapidity range 
$2<\eta<5$ is determined to be $\sigma_{\rm acc}= 62.2 \pm 0.2 \pm 2.5\mbarn$. 
The first uncertainty is the intrinsic systematic uncertainty of the measurement, 
the second is due to the uncertainty on the integrated luminosity. The statistical 
uncertainty is negligible. Extrapolation to full phase space yields the total 
inelastic proton-proton cross-section $\sigma_{\rm inel}= 75.4 \pm 3.0 \pm 4.5$\,mb,
where the first uncertainty is experimental and the second due to the extrapolation.
An updated value of the inelastic cross-section at a centre-of-mass energy of 7\,TeV
is also reported. 
\end{abstract}

\vspace*{2.0cm}

\begin{center}
  Published in JHEP 06 (2018) 100  
\end{center}

\vspace{\fill}

{\footnotesize 
\centerline{\copyright~CERN on behalf of the \lhcb collaboration, licence \href{http://creativecommons.org/licenses/by/4.0/}{CC-BY-4.0}.}}
\vspace*{2mm}

\end{titlepage}


\cleardoublepage


\renewcommand{\thefootnote}{\arabic{footnote}}
\setcounter{footnote}{0}

\cleardoublepage


\pagestyle{plain} 
\setcounter{page}{1}
\pagenumbering{arabic}



\section{Introduction}
\label{sec:Introduction}
The inelastic cross-section is a fundamental quantity in the phenomenology of
high-energy hadronic interactions that are studied at particle accelerators. It is 
also important for astroparticle physics, \eg\ in the description of extensive air 
showers induced by cosmic rays hitting the atmosphere of the Earth~\cite{PhysRevC.92.034906}, 
or for the modelling of the transport of cosmic ray particles in the interstellar 
medium~\cite{PhysRevD.90.085017,1475-7516-2015-09-023}. 
Since quantum chromodynamics cannot yet be solved in the nonperturbative regime, 
it is currently not possible to calculate the inelastic cross-section from 
first principles. Models based on Regge phenomenology predict, within the limits 
of the Froissart-Martin bound~\cite{PhysRev.123.1053,Martin1966}, an increase with 
energy according to a power law~\cite{landshoff2013}. Asymptotically the
Froissart-Martin bound grows proportional to $(\ln s)^2$, where $s$ is the square 
of the centre-of-mass energy of the collision. Although originally derived for the
total cross-section, this bound has been shown to apply also 
for the inelastic cross-section~\cite{PhysRevD.80.065013}. 

This paper presents a measurement of the inelastic proton-proton cross-section 
at $\sqrt{s}=13$\,TeV, which is the highest collision energy reached so far 
at any particle accelerator. The measurement is performed with the LHCb detector 
in the pseudorapidity range $2<\eta<5$. Other measurements of the inelastic 
proton-proton cross-section at LHC energies have been reported by the
ALICE~\cite{ALICE_xSecIn7tev} (2.76 and 7 TeV), 
ATLAS~\cite{ATLAS_xSecIn7tev, ATLAS_xSecTot7tev, ATLAS_xSecTot8tev, 
ATLAS_xSecIn13tev} (7, 8 and 13 TeV), 
CMS~\cite{CMS_xSecIn7tev, CMS_13TeV} (7 and 13 TeV),
LHCb~\cite{LHCb-PAPER-2014-057} (7 TeV) and
TOTEM~\cite{TOTEM_xSecTot7teV, TOTEM_xSecTot7teV_1, 
TOTEM_xSecTot7teV_2,TOTEM_xSecTot7tev_3,TOTEM_xSecTot8tev, TOTEM13TeV} (7, 8 and 13 TeV)
collaborations, covering also central and very forward rapidities.

\section{Detector and data samples}
\label{sec:Detector}
The \lhcb detector~\cite{Alves:2008zz,LHCb-DP-2014-002} is a single-arm forward
spectrometer, designed for 
the study of particles containing \bquark or \cquark quarks. The detector 
includes a high-precision tracking system consisting of a silicon-strip vertex 
detector surrounding the interaction region, a large-area silicon-strip 
detector located upstream of a dipole magnet with a bending power of about
$4{\mathrm{\,Tm}}$, and three stations of silicon-strip detectors and straw
drift tubes placed downstream of the magnet. The tracking system provides a 
measurement of momentum \ptot\ of charged particles with a relative uncertainty 
that varies from 0.5\% at low momentum to 1.0\% at 200\gevc. The minimum distance 
of a track to a primary vertex (PV), the impact parameter, is measured with 
a resolution of $(15+29/\pt)\mum$, where \pt is the component of the momentum 
transverse to the beam, in\,\gevc. Different types of charged hadrons are 
distinguished using information from two ring-imaging Cherenkov detectors. 
Photons, electrons and hadrons are identified by a calorimeter system 
consisting of scintillating-pad and preshower detectors, an electromagnetic
calorimeter and a hadronic calorimeter. Muons are identified by a system 
composed of alternating layers of iron and multiwire proportional chambers.

The online event selection for this measurement is based on unbiased triggers, 
which randomly accept a small subset of all bunch crossings. The bulk of the 
recorded data are from collisions between leading bunches in the bunch trains 
of the LHC filling pattern~\cite{LHC}, thus largely reducing background from 
previous bunch crossings. 
Data were collected for both polarities of the LHCb dipole magnet to 
test for magnetic-field dependent systematic effects. 
The total data sample consists of 691 million events in 49 
runs from 8 LHC fills, recorded in 2015 between July 8 and August 13. A run 
corresponds to a data set recorded under stable conditions and for a duration 
of up to one hour. Data from a long fill are spread over several runs.

The integrated luminosity of this data set was determined in a separate study.
The standard way to determine the relative luminosity in LHCb is
based on continuous monitoring of the rate of interactions with at least two 
tracks reconstructed in the vertex detector~\cite{LHCb-PAPER-2014-047}. 
This is done online by applying the empty-event counting method (see Sect. 3) 
to a dedicated set of randomly sampled events that are partially reconstructed 
in the trigger. The integrated luminosity is obtained by dividing the number 
of those interactions by their "reference" cross-section. With independent
data from a dedicated LHC fill at $\sqrt{s}=13$~TeV, this reference cross-section
was determined to be 63.4~mb with an uncertainty of 3.9\%, using the beam-gas 
imaging method as described in Ref.~\cite{LHCb-PAPER-2014-047}. For the unbiased data 
from leading bunch crossings the number of partially reconstructed events for the 
luminosity measurement is much smaller than the number of fully reconstructed events 
available for offline analysis. Therefore, to obtain precise relative luminosity 
measurements that permit sensitive studies of systematic effects, the empty-event 
counting method is applied to the fully reconstructed events. The analysis is 
performed per leading bunch crossing and in time intervals of $O(1{\rm s})$, 
thereby minimising systematic uncertainties due to differences in the individual 
bunch currents and variations of the instantaneous interaction rates. Differences 
between the partial reconstruction in the trigger and the full reconstruction 
result in a difference of about 1\% in the visible interaction rates. The ratio
was measured with a statistical uncertainty of 0.2\%. Accounting for this 
difference and taking the absolute calibration from the beam gas imaging method, 
a total integrated luminosity of 10.7~nb$^{-1}$ is obtained for the full data set, 
with an uncertainty of 4\%, which is dominated by the 3.9\%  uncertainty on the 
reference cross-section. Additional contributions are the 0.2\% statistical 
uncertainty of the cross-calibration factor and a 0.8\% difference when requiring 
at least one reconstructed primary vertex instead of two vertex-detector tracks.

Simulated events are used to study the detector response and effects of the 
reconstruction chain. In the simulation, proton-proton collisions for both magnet 
polarities are generated using 
\mbox{\pythia$\!\!$\,8}~\cite{Sjostrand:2007gs,Sjostrand:2006za} 
with a specific \lhcb configuration~\cite{LHCb-PROC-2010-056}. 
Decays of hadronic particles are described by \evtgen~\cite{Lange:2001uf}, 
in which final-state radiation is generated using \photos~\cite{Golonka:2005pn}. 
The interaction of the generated particles with the detector, and its response,
are implemented using the \geant toolkit~\cite{Allison:2006ve, *Agostinelli:2002hh} 
as described in Ref.~\cite{LHCb-PROC-2011-006}.

\section{Analysis method}
\label{sec:strategy}
The primary measurement is a fiducial cross-section, defined as the cross-section 
for proton-proton collisions with at least one prompt, long-lived charged particle 
with momentum $p>2$\,GeV/$c$ and pseudorapidity in the range $2<\eta<5$. A particle
is defined as ``long-lived'' if its lifetime is larger than 30\,ps, and it is prompt 
if it is produced directly in the primary collision or if none of its ancestors is 
long-lived. At the LHCb experiment a lifetime of 30\,ps corresponds to a typical 
flight length of $O(100)$\,mm. According to this definition, for instance, ground-state 
hyperons are long-lived, but not any particle containing charm or beauty quarks. 

The experimental selection of prompt long-lived charged particles requires well 
reconstructed charged tracks with momentum $p>2$\,GeV/$c$ and $2<\eta<5$ that traverse 
the entire LHCb tracking system and have an estimated point of origin located 
longitudinally (along the beam direction) within 200\,mm and transversally within 
0.4\,mm of the average PV position in the run. From a parametrisation of the 
PV density by a three-dimensional Gaussian function, the estimated point of origin 
is determined as that point on the particle trajectory, parametrised by a straight 
line, where the PV density is highest. With this selection all events can be used 
in the analysis, independently of whether a PV was reconstructed. The above 
requirements select almost exclusively inelastic interactions. From about 8.7 million
elastic proton-proton scattering processes in the simulation none is accepted. 

The cross-section measurement exploits the fact that the recorded event sample is 
unbiased, with the number of inelastic interactions per event drawn from a Poisson 
distribution. The average number of interactions $\mu$ per event can then be 
inferred from the fraction $p_0$ of empty events, $\mu=-\ln p_0$, and for a given 
number $N_{\rm evt}$ of events the fiducial cross-section is given by
\beq{1} 
\sigma_{\rm acc}=\frac{(\mu - \mu_{\rm bkg})N_{\rm evt}}{\lum} \;,
\eeq
where $\lum$ is the integrated luminosity of the event sample. The number 
$\mu_{\rm bkg}$ of background interactions per event is estimated from bunch 
crossings where only the bunch from one of the beams was populated. 
The largest background levels are found for the first LHC fill
used in the analysis, with $\mu_{\rm bkg}/\mu$ around 1\%.
The cross-section measurement is performed separately for all leading bunch 
crossings, and in time intervals of $O(8{\rm s})$ to follow variations of the 
interaction rate during a run.   

The determination of the empty-event probability $p_0$ takes into account 
that, because of inefficiencies, events may be wrongly tagged as empty,
and that events which have no prompt long-lived charged particle inside the 
fiducial region can be classified as non-empty because of misreconstructed tracks. 
For the measurement presented here, the detector related effects are accounted 
for by an approach that relates $p_0$ to the observed charged track multiplicity 
distribution inside the fiducial region.

A good approximation for the low-multiplicity events that dominate the empty-event 
counting is the assumption that on average the detector response is the same for 
every true particle. In other words, the multiplicity distribution of reconstructed 
tracks is assumed to be the same for every true particle. As shown below, in this 
case $p_0$ can be determined from the observed multiplicity distribution of 
long-lived prompt charged tracks in the detector acceptance. 

The relation between $p_0$ and experimentally accessible information 
can be derived starting from the probability generating function 
(PGF) of the observed multiplicity distribution $F_q(x)= \sum_n q_n x^n$, where 
the probability $q_n$ to observe $n$ tracks is weighted by the $n$-th power 
of a continuous variable $x$. It can be shown that the PGF of a convolution 
of two discrete probability distributions is the product of the individual 
PGFs. Introducing $G(x)$ as the PGF of the multiplicity distribution that is 
reconstructed for a single true particle, the PGF for the case of $k$ true
particles is the PGF of the convolution of $k$ single particle distributions,
\ie\ the $k$-th power $G^k(x)$. 
Weighting each true multiplicity $k$ with its probability $p_k$, 
the relation between the PGF of the observed multiplicity 
distribution $q_n$ and the true multiplicity distribution $p_k$ is given by
\beq{2}
     F_q(x) = \sum_{n=0}^\infty q_n x^n = \sum_{k=0}^\infty p_k\,G^k(x)  \;.
\eeq
The true empty-event probability $p_0$ can be inferred by setting
$x=\alpha$ such that $G(\alpha)=0$, which yields
\beq{3}
    p_0 = \sum_{n=0}^\infty q_n \alpha^n \;.
\eeq

The parameter $\alpha$ is the only detector-related parameter of the analysis.
It is an unfolding parameter that relates $p_0$ to the observed charged particle 
multiplicity distribution in the fiducial region. For an ideal detector it would be 
zero. For a given experiment the value of $\alpha$ depends mainly on the average 
reconstruction efficiency. Assuming for example a binomial detector response, where 
a particle is either reconstructed with efficiency $\eps$ or missed, one has 
$G(x)=(1-\eps) +\eps x$ and thus $\alpha=(\eps-1)/\eps$, which is always negative. 
When taking $p_0$ and $q_n$ from fully simulated events and solving Eq.\,(\ref{3}) 
for $\alpha$, one obtains an effective parameter that also accounts for 
higher-order effects due to background tracks and nonlinear detector response. 

For proton-proton collisions at high centre-of-mass energies, where inelastic 
interactions have high multiplicity final states, and for data with a small average 
number of simultaneous interactions per bunch crossing, the cross-section measurement
has only very little sensitivity to the exact value of $\alpha$. The measurements  
presented below are based on events with $\mu$ in the range between 0.4 and 1.4
and values of $q_0$ that are at least an order of magnitude larger than 
the values $q_n$ for $n>0$. With a typical value $\alpha \approx -0.6$ the 
values of $p_0$ are on average only about 3\% smaller than their leading-order 
estimates $q_0$, which results in robust cross-section measurements even in 
case of sizeable systematic uncertainties on $\alpha$.

\section{Measurement of the fiducial cross-section}
The inelastic fiducial cross-section is determined separately for all runs recorded
with unbiased triggers and, within a run, all leading bunch crossings. In total 243 
independent measurements are done, with different filling patterns of the LHC, 
different bunch currents and both magnet polarities. For each measurement an
initial estimate for the unfolding parameter $\alpha$ is obtained from a simulation 
that has been weighted to match the average reconstructed track multiplicity in data. 
This initial value is then corrected to account for differences between data and 
simulation in the average track reconstruction efficiency and the average fraction 
of misreconstructed tracks. The efficiency correction uses an independent calibration 
for the analysed data set, determined as described in Ref.~\cite{LHCb-DP-2013-002}. 
The fraction of misreconstructed tracks is estimated from the fraction of tracks 
rejected by the track selection criteria, with a constant of proportionality taken 
from simulation. The observed differences between data and simulation are propagated 
into $\alpha$ by means of a simplified model that relates it to the average 
track reconstruction efficiency and the fraction of misreconstructed tracks. 

The individual cross-section measurements are combined in a weighted average, 
assuming uncorrelated statistical and fully correlated systematic uncertainties.
The weight of each measurement is proportional to the integrated luminosity
of the corresponding data set, resulting in an overall fiducial cross-section 
$\sigma_{\rm acc} = 62.237 \pm 0.002$\,mb, where the uncertainty is purely 
statistical. The contributions to the systematic uncertainty are summarised in 
Table\,\ref{tab:syst}. The dominant contribution is the 4\% uncertainty on the 
integrated luminosity. The intrinsic uncertainty of the analysis is driven by a
16\% uncertainty on the unfolding parameter $\alpha$, which propagates into a 
0.25\% systematic uncertainty on $\sigma_{\rm acc}$. The largest contribution is 
due to the difference between either determining $\alpha$ from all simulated events 
or only from events with particles inside the fiducial region. The systematic 
uncertainties due to the efficiency calibration and the differences in the fraction 
of misreconstructed tracks between data and simulation, where the full size of the 
correction is assigned as a systematic uncertainty, are slightly smaller.

\begin{figure}[tb]
\centering
\includegraphics[width=0.8\textwidth]{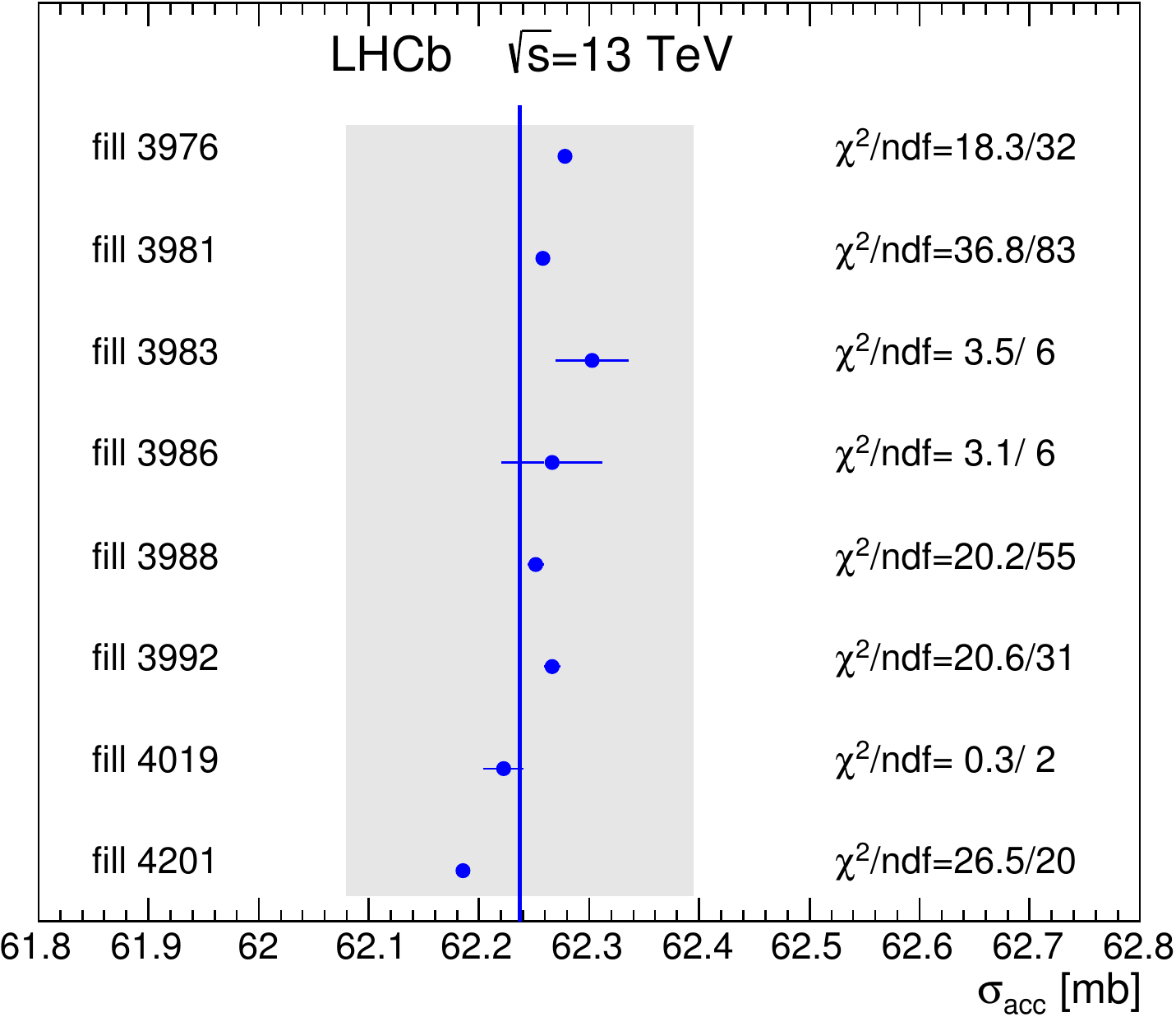}
\caption{\label{fig:xsecfill}\small 
Overall fiducial cross-section (vertical line), compared to the averages of
the individual results in different LHC fills. The error bars indicate the 
statistical uncertainties. The grey band indicates the systematic uncertainty 
on the overall average due to the unfolding parameter $\alpha$. The $\chi^2$-values 
for the averages inside a fill are calculated with only the statistical 
uncertainties and the number of degrees of freedom ($\rm ndf$) is one less than 
the number of individual results contributing to the average. Systematic 
uncertainties inferred from the observed spread between the fills are 
discussed in the text.}
\end{figure}

Figure\,\ref{fig:xsecfill} shows a comparison of the overall fiducial cross-section 
with the averages within the individual LHC fills. While within a fill all 
measurements are found to be consistent within their statistical uncertainties, 
small but significant differences are seen between fills. These differences
are found to be correlated with quantities not studied in the simulation, namely 
the vertical position and extension of the luminous region and, to a lesser extent, 
the background level seen in the data. The spread associated to those variables 
corresponds to an additional systematic uncertainty of $0.05\%$. Also given in
Fig.\,\ref{fig:xsecfill} are the $\chi^2$-values of the individual averages, 
calculated with only the statistical uncertainties. Inspection of the $\chi^2$-values 
shows that, except for the last fill, the agreement between the results within one 
fill is actually better than expected. This is due to the fact that the 
luminosity calibration and the inelastic cross-section measurement 
are correlated by the use of information recorded by the vertex detector. 
The average for the last 
fill, which in comparison to the others has an enlarged $\chi^2$ value, is 
dominated by two runs with more than 100 million events. This points to the 
existence of additional systematic effects of about the size of the statistical 
uncertainty of this average, which in view of the other uncertainties are 
negligible. Cross-checks from variations of the track selection 
criteria show no indication of additional systematic effects.
 
\begin{table}[htb]
\centering
\caption{\label{tab:syst}\small
Summary of systematic uncertainties on the fiducial cross-section.
For the contribution from the unfolding parameter $\alpha$ a breakdown
into the individual components is given.}
\begin{tabular}{lc}
  \multicolumn{1}{l}{Source}  & Relative uncertainty \\
\hline
  Integrated luminosity              &     4.00\%  \\
  Unfolding parameter $\alpha$       &     0.25\%  \\
  --- Interactions not in acceptance & \hspace{10mm}  0.18\%  \\
  --- Efficiency                     & \hspace{10mm}  0.15\%  \\
  --- Misreconstructed tracks        & \hspace{10mm}  0.12\%  \\
  Luminous region and background     &     0.05\%  \\ 
\hline
  Total                              &     4.01\% 
\end{tabular}
\end{table}

\section{Extrapolation to full phase space}
The extrapolation from the fiducial cross-section $\sigma_{\rm acc}$ to the total 
inelastic cross-section $\sigma_{\rm inel} = F_{\rm T} \,\sigma_{\rm acc}$ follows the same 
approach as in Ref.~\cite{LHCb-PAPER-2014-057}. The extrapolation factor $F_{\rm T}$
is determined from generator-level simulations. Neglecting interference effects 
between different contributions, it is assumed that the total inelastic 
cross-section can be written as an incoherent sum of distinct contributions 
\beq{4}
    \sigma_{\rm inel} = \sum_X \sigma_X
    \quad\mbox{with}\quad
     X \in \{\rm ND, SDA, SDB, DD\} \;.
\eeq
Here $\sigma_{\rm ND}$ is the non-diffractive cross-section, $\sigma_{\rm SDA}$ and 
$\sigma_{\rm SDB}$ are the single diffractive contributions with the diffractively
excited system travelling towards (A) or away (B) from the detector, which have the 
same cross-section but different contributions to the visible cross-section, 
and $\sigma_{\rm DD}$ is the double diffractive cross-section. State-of-the-art 
event generators are assumed to provide a realistic parametrisation of the properties 
of the various contributions. This has been studied with the 32 proton-proton tunes 
that come with \pythia\,8.230~\cite{sjostrand:2015xx} and which do not require 
external libraries. Table\,\ref{tab:tunes} gives mean 
values and standard deviations of the fractions $f_X$ of the inelastic cross-section, 
the fractions $v_X$ of interactions with at least one prompt long-lived charged 
particle within the acceptance and, for those interactions, the average 
multiplicities $n_{{\rm ch},X}$ of those particles inside the acceptance. 

\begin{table}[tb]
\centering
\caption{\label{tab:tunes}\small
Properties of \pythia{}\,8.230 proton-proton tunes. Mean values and standard deviations 
are given for the fractions $f_X$ of the inelastic cross-section, the fractions 
$v_X$ of interactions inside the acceptance and, for those interactions, the 
average numbers of long-lived prompt charged particles $n_{{\rm ch},X}$ inside 
the acceptance.}
\begin{tabular}{lrrrrrr}
 & \multicolumn{2}{c}{$f_X$}
 & \multicolumn{2}{c}{$v_X$}
 & \multicolumn{2}{c}{$n_{{\rm ch},X}$}\\[1mm]
 &  mean & \multicolumn{1}{c}{s.d.} 
 &  mean & \multicolumn{1}{c}{s.d.} 
 &  mean & \multicolumn{1}{c}{s.d.} \\
\hline
Non-diffractive    (ND)  & 0.720 & 0.012 & 0.9963 & 0.0005 & 17.94 & 1.45 \\
Single diffractive (SDA) & 0.083 & 0.003 & 0.7154 & 0.0051 &  8.11 & 0.52 \\
Single diffractive (SDB) & 0.083 & 0.003 & 0.3411 & 0.0077 &  7.83 & 0.44 \\
Double diffractive (DD)  & 0.114 & 0.006 & 0.6263 & 0.0049 &  6.15 & 0.31 \\
\end{tabular}
\end{table}

Given the fractions $f_X$ of the total inelastic cross-section and the fractions 
of visible interactions $v_X$, the extrapolation factor $F_{\rm T}$ is
\beq{5}
  F_{\rm T} = \frac{\sum_X \sigma_X}{\sum_X \sigma_X\,v_X}
      = \frac{1}{\sum_X f_X \, v_X} \;.
\eeq
Taking the standard deviations from Table\,\ref{tab:tunes} as model uncertainties 
would likely underestimate the uncertainty of the extrapolation factor, 
since in particular the cross-section fractions have a much smaller spread 
than the uncertainties obtained in a measurement of the diffractive contributions 
to the inelastic cross-section, $f_{\rm SD}=0.20^{+0.04}_{-0.07}$ and $f_{\rm DD}=0.12^{+0.05}_{-0.04}$, 
performed by the ALICE collaboration at $\sqrt{s}=7$\,TeV~\cite{ALICE_xSecIn7tev}. 

To reduce the model dependence in the determination of $F_{\rm T}$, the cross-section 
fractions are considered to be a priori unknown and only subject to the constraint 
$\sum_X f_X=1$. The extrapolation factor is estimated from sets $\{f_X\}$ that 
uniformly sample the subspace defined by this constraint. For each set $\{f_X\}$ 
the extrapolation factor $F_{\rm T}$ and the average multiplicity 
$n_{\rm ch} = \sum_X f_X\,n_{{\rm ch},X}$ inside the fiducial region are 
calculated using $v_X$ and $n_{{\rm ch},X}$ from Table~\ref{tab:tunes}.
The spread of the different tunes is propagated into the extrapolation
factor by drawing $v_X$ and $n_{{\rm ch},X}$ from Gaussian distributions with 
mean values and standard deviations as given in the table.
An additional experimental constraint is imposed by assigning a Gaussian 
weight $w=\exp(-(n_{\rm ch}-N)^2/2\sigma^2_N)$ to $\{f_X\}$ and $F_{\rm T}$, where 
$N=13.9 \pm 0.9$ is the average multiplicity per interaction of prompt long-lived
charged particles inside the acceptance in the data. The numerical value for this
constraint is obtained from the full simulation, tuned to reproduce the observed 
average multiplicity per event and corrected for differences between data and 
simulation in the average track reconstruction efficiency and the fraction of tracks 
that are associated to a true particle.

\begin{figure}[tb]
\centering
\includegraphics[width=0.9\textwidth]{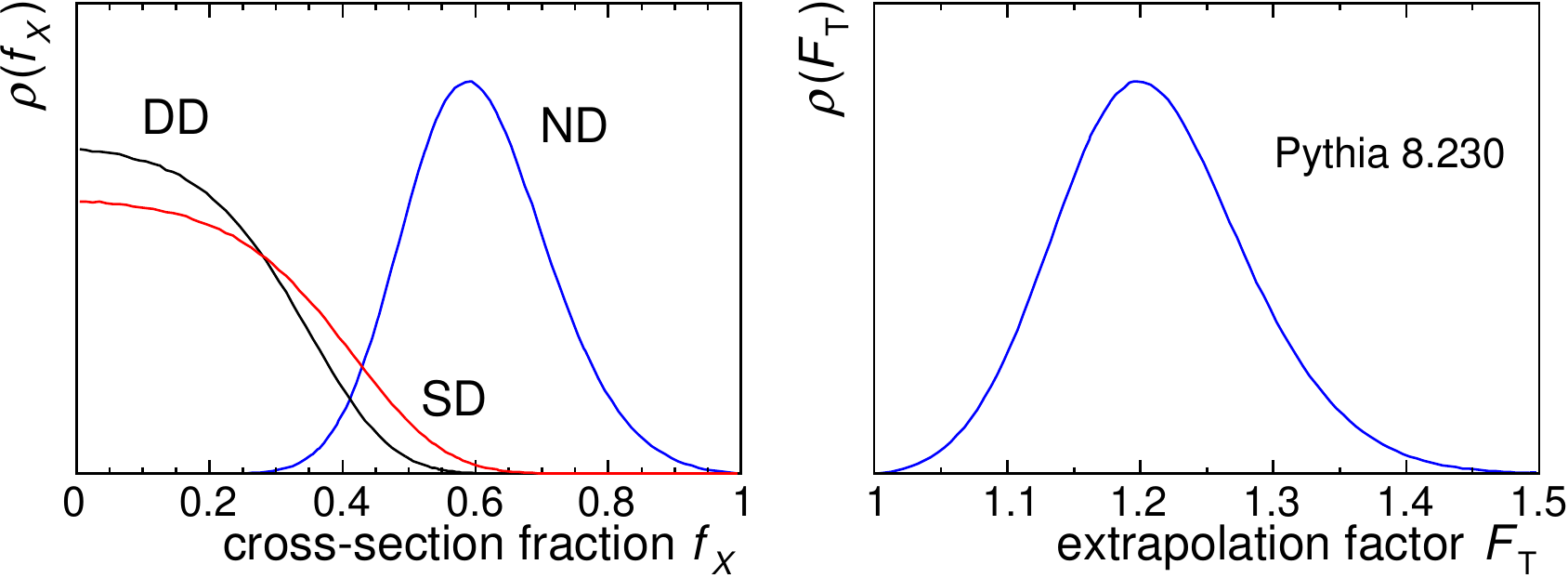}
\caption{\label{fig:scal}\small
Posterior densities of (left) the cross-section fractions $f_X$ for non-diffractive (ND) 
double-diffractive (DD) and single-diffractive (SD=SDA+SDB) contributions, and (right) 
of the extrapolation factor $F_{\rm T}$.}
\end{figure}

Figure\,\ref{fig:scal} shows the posterior densities $\rho(f_X)$ and $\rho(F_{\rm T})$ 
of the cross-section fractions $f_X$ and the cross-section extrapolation factor 
$F_{\rm T}$. The mean values of the fractions of $f_X$ are found to be 
$f^{\rm sim}_{\rm SD}=0.21$ and $f^{\rm sim}_{\rm DD}=0.18$, consistent with 
measurements at $\sqrt{s}=7$\,TeV~\cite{ALICE_xSecIn7tev}. The resulting 
cross-section extrapolation factor is $F_{\rm T} = 1.211 \pm 0.072$, which yields 
a total inelastic cross-section of 
\begin{equation*}
   \sigma_{\rm inel} = 75.4 \pm 3.0(\rm exp) \pm 4.5(\rm extr) \,{\rm mb} \;,
\end{equation*}
where the first uncertainty is due to the experimental uncertainty of the
fiducial cross-section and the second due to the cross-section extrapolation. 
Summing all uncertainties in quadrature one finds $\sigma_{\rm inel} = 75.4 \pm 5.4$\,mb.

\section{Summary and conclusions}
A measurement is presented of the inelastic proton-proton cross-section with 
at least one prompt long-lived charged particle with momentum $p>2$\,GeV/$c$ 
in the pseudorapidity range $2<\eta<5$. A particle is defined as ``long-lived'' 
if its lifetime is larger than 30\,ps, and it is prompt if it is produced 
directly in the primary interaction or if none of its ancestors is long-lived. 
The measurement is done with the empty-event counting method applied to unbiased 
data. A total of 691 million events is analysed. The statistical 
uncertainty of the overall result is negligible. The systematic uncertainty
has contributions from the integrated luminosity (4\%), the unfolding
parameter (0.25\%) and vertical location and extension of the luminous
region and background levels (0.05\%). Adding all uncertainties not related 
to the integrated luminosity in quadrature, the final result for the 
fiducial cross-section is
\begin{equation*}
   \sigma_{\rm acc}(\sqrt{s}=13\,{\rm TeV}) 
  = 62.2 \pm 0.2 \pm 2.5(\rm lumi) \,{\rm mb}  \;.
\end{equation*}
Extrapolating to the full phase space yields a total inelastic cross-section of
\begin{equation*}
   \sigma_{\rm inel}(\sqrt{s}=13\,{\rm TeV})  
 = 75.4 \pm 3.0(\rm exp) \pm 4.5(\rm extr) \,{\rm mb} \;. 
\end{equation*}

Since the publication of a measurement of the inelastic proton-proton 
cross-section at a centre-of-mass energy of 7\,TeV by the LHCb 
collaboration~\cite{LHCb-PAPER-2014-057} an improved calibration of the 
luminosity scale has become available~\cite{LHCb-PAPER-2014-047}. The new 
value of the reference cross-section for the integrated luminosity 
of the data analysed for the previous measurement is 2.7\% larger than the 
initial estimate and the uncertainty has been reduced from 3.5\% to 1.7\%. 
With the analysis of Ref.~\cite{LHCb-PAPER-2014-057} unchanged, the updated 
cross-section is
\begin{equation*} 
   \sigma_{\rm inel}(\sqrt{s}=7\,{\rm TeV}) 
  =68.7 \pm 2.1(\rm exp) \pm  4.5(\rm extr)\,{\rm mb}\;,
\end{equation*}
which supersedes the previous result. The experimental uncertainty is reduced 
from 4.3\% to 3.0\% and the central value shifted up by 2.7\%. 

\begin{figure}[tb]
\centering
\includegraphics[width=0.925\textwidth]{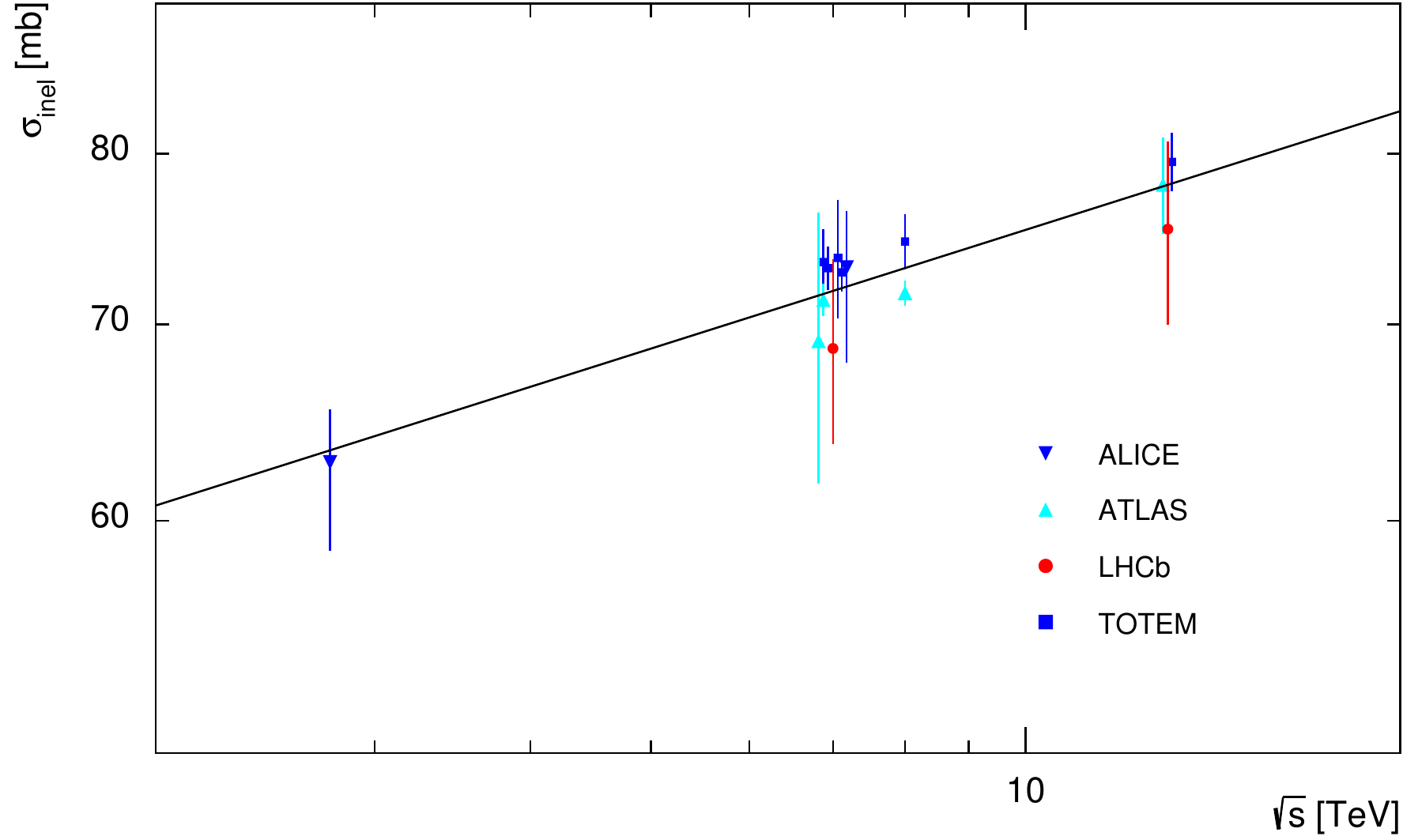}
\caption{\label{fig:overview}\small 
Measurement of the total inelastic proton-proton cross-section at the LHC
at centre-of-mass energies of 2.76, 7, 8 and 13 TeV. Results are shown
from the 
ALICE~\cite{ALICE_xSecIn7tev}, 
ATLAS~\cite{ATLAS_xSecIn7tev, ATLAS_xSecTot7tev, ATLAS_xSecTot8tev, 
ATLAS_xSecIn13tev} and TOTEM~\cite{TOTEM_xSecTot7teV, TOTEM_xSecTot7teV_1, 
TOTEM_xSecTot7teV_2,TOTEM_xSecTot7tev_3,TOTEM_xSecTot8tev, TOTEM13TeV} collaborations.
For better visibility, measurements at the same energy are drawn at slightly 
displaced locations. The error bars show the total uncertainties, with 
positive and negative uncertainties of the respective results independently 
added in quadrature. The line shows the result from a power-law fit.}
\end{figure}

A comparison of the total inelastic cross-section measurements from  proton-proton 
collisions at the LHC is shown in Fig.~\ref{fig:overview}. The new LHCb 
measurement at $\sqrt{s}=13$\,TeV is in good agreement with the 
measurements by the ATLAS~\cite{ATLAS_xSecIn13tev} and TOTEM~\cite{TOTEM13TeV} 
collaborations. In the LHC energy range the dependence of the inelastic 
cross-section on $\sqrt{s}$ is well described by a power law. 

\section*{Acknowledgements}
%
%
\noindent We express our gratitude to our colleagues in the CERN
accelerator departments for the excellent performance of the LHC. We
thank the technical and administrative staff at the LHCb
institutes. We acknowledge support from CERN and from the national
agencies: CAPES, CNPq, FAPERJ and FINEP (Brazil); MOST and NSFC
(China); CNRS/IN2P3 (France); BMBF, DFG and MPG (Germany); INFN
(Italy); NWO (The Netherlands); MNiSW and NCN (Poland); MEN/IFA
(Romania); MinES and FASO (Russia); MinECo (Spain); SNSF and SER
(Switzerland); NASU (Ukraine); STFC (United Kingdom); NSF (USA).  We
acknowledge the computing resources that are provided by CERN, IN2P3
(France), KIT and DESY (Germany), INFN (Italy), SURF (The
Netherlands), PIC (Spain), GridPP (United Kingdom), RRCKI and Yandex
LLC (Russia), CSCS (Switzerland), IFIN-HH (Romania), CBPF (Brazil),
PL-GRID (Poland) and OSC (USA). We are indebted to the communities
behind the multiple open-source software packages on which we depend.
Individual groups or members have received support from AvH Foundation
(Germany), EPLANET, Marie Sk\l{}odowska-Curie Actions and ERC
(European Union), ANR, Labex P2IO and OCEVU, and R\'{e}gion
Auvergne-Rh\^{o}ne-Alpes (France), Key Research Program of Frontier
Sciences of CAS, CAS PIFI, and the Thousand Talents Program (China),
RFBR, RSF and Yandex LLC (Russia), GVA, XuntaGal and GENCAT (Spain),
Herchel Smith Fund, the Royal Society, the English-Speaking Union and
the Leverhulme Trust (United Kingdom).


\clearpage
\addcontentsline{toc}{section}{References}
\bibliographystyle{LHCb}
\bibliography{main,LHCb-PAPER,LHCb-DP}

\newpage
\centerline{\large\bf LHCb collaboration}
\begin{flushleft}
\small
R.~Aaij$^{43}$,
B.~Adeva$^{39}$,
M.~Adinolfi$^{48}$,
Z.~Ajaltouni$^{5}$,
S.~Akar$^{59}$,
J.~Albrecht$^{10}$,
F.~Alessio$^{40}$,
M.~Alexander$^{53}$,
A.~Alfonso~Albero$^{38}$,
S.~Ali$^{43}$
% ...

{\footnotesize \it
% ...
$ ^{5}$Clermont Universit{\'e}, Universit{\'e} Blaise Pascal, CNRS/IN2P3, LPC, Clermont-Ferrand, France\\
% ...
$ ^{10}$Fakult{\"a}t Physik, Technische Universit{\"a}t Dortmund, Dortmund, Germany\\
% ...
$ ^{38}$ICCUB, Universitat de Barcelona, Barcelona, Spain\\
$ ^{39}$Instituto Galego de F{\'\i}sica de Altas Enerx{\'\i}as (IGFAE), Universidade de Santiago de Compostela, Santiago de Compostela, Spain\\
$ ^{40}$European Organization for Nuclear Research (CERN), Geneva, Switzerland\\
% ...
$ ^{43}$Nikhef National Institute for Subatomic Physics, Amsterdam, The Netherlands\\
% ...
$ ^{48}$H.H. Wills Physics Laboratory, University of Bristol, Bristol, United Kingdom\\
% ...
$ ^{53}$School of Physics and Astronomy, University of Glasgow, Glasgow, United Kingdom\\
% ...
$ ^{59}$University of Cincinnati, Cincinnati, OH, United States\\
% ...
}
\end{flushleft} 
\end{document}